\documentstyle[prl,aps,multicol,epsf]{revtex}

\def\mathrm#1{\mbox{#1}}
\def\approxeq{\approx}
\newcommand{\htc}{high-T$_{\mathrm{c}}$ }
\newcommand{\lasrco}{La$_{2-x}$Sr$_x$CuO$_4$ }
\newcommand{\ybco}{YBa$_2$Cu$_3$O$_{7-x}$ }
\newcommand{\cacocl}{Ca$_{2}$CuO$_{2}$Cl$_{2}$ }
\newcommand{\tp}{t^\prime}
\newcommand{\tpp}{t^{\prime\prime}}
\newcommand{\INT}{\mathrm{int}}
\newcommand{\kin}{\mathrm{kin}}
\newcommand{\neel}{N\'{e}el }
\newcommand{\eps}{\varepsilon}
\newcommand{\af}{\mathrm{AF}}
\newcommand{\SC}{\mathrm{SC}}
\newcommand{\mod}{\mathrm{mod}}
\newcommand{\sgn}{\mathrm{sgn}}
\newcommand{\up}{\uparrow}
\newcommand{\down}{\downarrow}
\newcommand{\ul}{}
\begin{document}
\draft

\title{Interrelation of Superconducting and Antiferromagnetic Gaps in \htc
Compounds: a Test Case for a Microscopic Theory}

\author{Werner Hanke$^{1,2}$, Marc G.\ Zacher$^1$, Enrico Arrigoni$^1$, Shou-Cheng Zhang$^2$}
\address{$^1$Institute for Theoretical Physics, University of  W\"urzburg,
97074 W\"urzburg, Germany\\
$^2$ Department of Physics, Stanford University, Stanford, California 94305, USA
}

\maketitle

\begin{abstract}
Recent angle resolved photoemission (ARPES) data, which found evidence 
for a $d$-wave-like modulation of the
antiferromagnetic gap, suggest an intimate interrelation between the 
antiferromagnetic insulator and
the superconductor with its $d$-wave gap. This poses a new challenge to microscopic
descriptions, which should account for this correlation between, at first sight, very
different states of matter. 
Here, we propose a microscopic mechanism which provides a definite correlation between these
two different gap structures: it is shown that a projected SO(5) theory, 
which aims at unifying antiferromagnetism and $d$-wave superconductivity
via a common symmetry principle while explicitly taking the Mott-Hubbard gap into account, 
correctly describes the observed gap characteristics.
Specifically, it
accounts for both the dispersion and the order of magnitude difference
between the antiferromagnetic gap modulation and the superconducting gap.
\end{abstract}

\begin{multicols}{2}

Especially due to recent advances in experimental techniques, such as
angular-resolved photoemission spectroscopy (ARPES) 
\cite{ronning,shenloeserwells,marshall,ding,kim,ino}
and
spectroscopy probing magnetic correlations \cite{aepplikastner,mook}, a crucial
basic ingredient seems to emerge for the phenomenology of the \htc
compounds:
in contrast to \ul{many} of the earlier
theoretical studies, which started from the highly doped system far away
from the antiferromagnetic (AF) insulator, the appropriate starting point
seems to be the insulating state, corresponding to half-filling in
theoretical models or to the undoped situation in experiment. A prominent
recent example, which advocates this point of view strongly, is a
photoemission observation by the Stanford group \cite{ronning}. It found evidence
suggesting a direct correlation between the $d-$wave symmetry of the
superconducting (SC) gap and an observed $d-$wave-like modulation of the AF gap.
Further experiments employing magnetic spectroscopy, support this view of an
intimate interrelation of the AF insulator and the SC state as well \cite{aepplikastner}. 
These
experimental developments have given the \htc research a new focus
and challenge: any microscopic theory has to account for this interrelation
between, at first glance, radically different states.

\begin{figure}
\epsfxsize=8cm
\epsffile{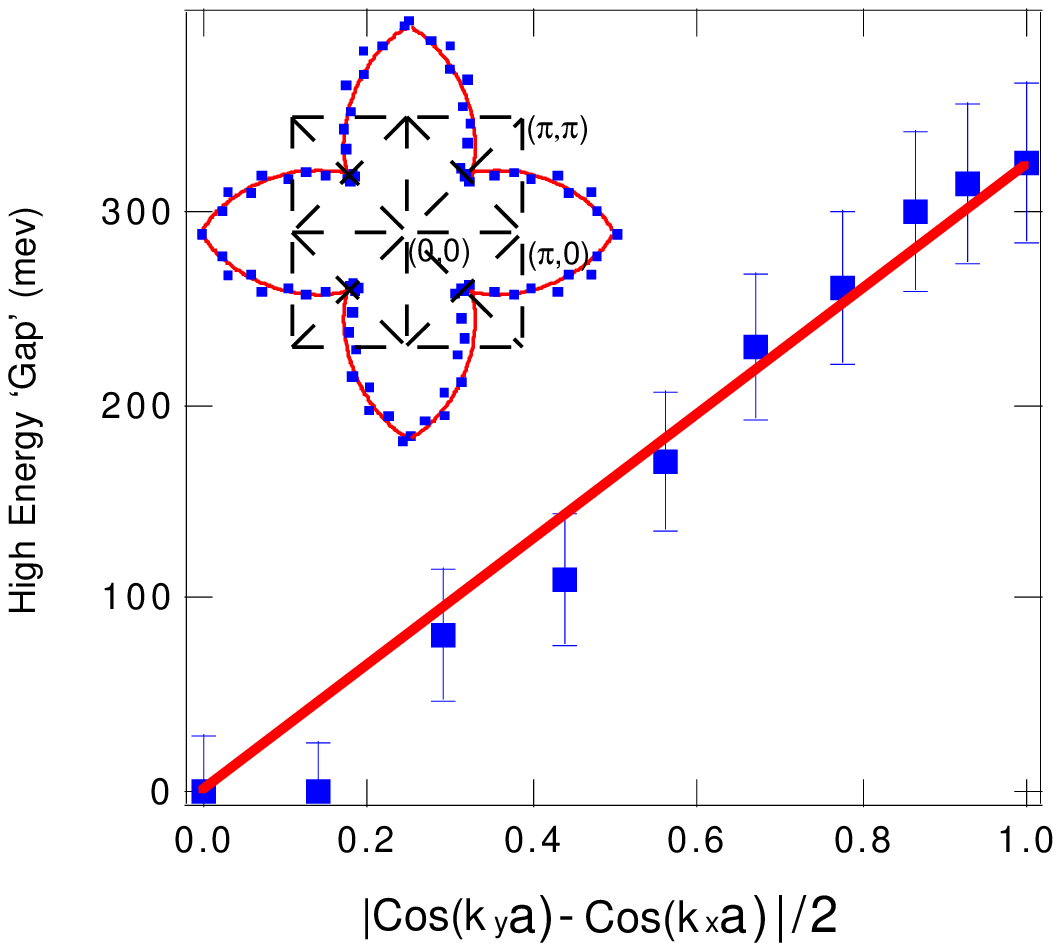}
\caption{ARPES results for Ca$_2$CuO$_2$Cl$_2$: 
The lowest energy peak disperses away from the $(\pi/2,\pi/2)$ peak position as one moves
away from wavevector $(\pi/2,\pi/2)$. The difference (gap-modulation) between this dispersive 
peak and the
$(\pi/2,\pi/2)$ peak is plotted against $|\cos(k_x a) - \cos(k_y a)|$. The data fit the
straight line which indicates a $d$-wave like dispersion.
The inset shows the gap modulation in momentum space along the antiferromagnetic zone
boundary. By plotting only the {\it difference} to $(\pi/2,\pi/2)$, the plot goes to
zero for this point by definition.}
\label{figshen}
\end{figure}

The ''gap'' structure in the AF phase as found by ARPES experiments in
insulating \cacocl \cite{ronning} is summarized in Fig. \ref{figshen}. These
ARPES data display a $d-$wave-like, i.e. $|\cos k_{x}-\cos k_{y}|$-like
dispersion in the one-electron spectral function $A\left( 
\vec k
,\omega \right) $ with respect to the lowest energy state at $\left( \frac{%
\pi }{2},\frac{\pi }{2}\right) $. The straight line in Fig. \ref{figshen} shows the $d-$%
wave dispersion function along the edge of the magnetic Brillouin Zone (dashed square in
the inset of Fig. \ref{figshen}) with a $d-$wave ''energy gap''. 
The inset of Fig. \ref{figshen} presents the ARPES data and
the ''gap'' features in a two-dimensional plot: on a line drawn from the
center of the Brillouin zone to the experimental points, the distance
between these points to the intersection of this line with the AF Brillouin
zone gives the value of the ''gap'' at the $\vec k-$point
considered. The data closely follow the $|\cos k_{x}-\cos k_{y}|$%
-dispersion, depicted in the $d-$wave full line. This ``$d$-wave" like
gap is a modulation of the uniform ($s$-wave) Mott-Hubbard gap of the order of
$U\sim eV$ of the insulating state. Therefore, throughout this paper,
we shall use the notation $s+|d|$ to characterize the full gap found 
in the AF insulator. The crucial point is that
these photoemission data suggest that the $|d|$ component of the AF gap in the
insulator is also the underlying reason for the celebrated pseudo-gap in the
underdoped regime: this ''high-energy'' pseudo-gap of the order $J\sim 0.1eV$
continuously evolves out of the insulating feature, as documented not only
by the same energy scale but again by the same $d-$wave dispersion \cite{marshall}.
Since, on the other hand, this high-energy feature is
closely correlated to the superconducting gap 
as a function of both doping and momentum 
\cite{marshall,ding,white},
we finally arrive at a crucial constraint on the
microscopic theory: such a theory should be able to explain the
interrelation between the superconducting gap and the AF gap modulation.
Theoretical concepts for the pseudo-gap range from descriptions based on
preformed pairs \cite{sevex}, schemes employing resonating valence bond (RVB)
singlet formation 
\cite{Tanamoto,wen}
and spin-charge separations \cite{laughlin} to
damped spin-density waves \cite{schmalian}. However, the microscopic origin of the
pseudo-gap and, in particular, its interrelation with the insulating gap
features are still unestablished.

In this report, we shall argue that there exists a theory, which provides
rather naturally such an interrelation. This is the SO(5) theory of
high-temperature superconductivity which aims at unifying AF and SC via a
symmetry principle 
\cite{zhang,meixner,ederetal98,rabello,henley,ederetal99,demler98,arrigoni99}.
The symmetry enables the construction of a five--dimensional ''superspin'' 
composed out of the three--component AF order parameter and the two--component SC order
parameter. An SO(5) ''rotation'' maps the AF microscopic state 
and its order parameter
onto
the superconducting counterpart and vice versa and, thereby, provides a
definite prediction of the gap characteristics of both states of matter:
we shall demonstrate that a $d-$wave
SC gap and the $|d|$ component of the AF gap arise from the same physical
origin and can be obtained from each other by an appropriate SO(5) rotation.
The obtained AF gap modulation has a characteristic sharp cusp around its
minimum at $\left( k_{x},k_{y}\right) =\left( \pm \frac{\pi }{2},\pm \frac{%
\pi }{2}\right) $, consistent with the available experimental data \cite{notedisp}. 
Furthermore,
the size of the AF gap modulation and the SC gap are also in good agreement 
with the experimental values.


Besides finding a modulation of the gap in the AF insulator, the recent
ARPES experiment\cite{ronning} also found a remnant Fermi surface of the
AF insulator. This shows that it is appropriate to think of the AF insulator
in terms of a condensate of {\it magnons} on top of a Fermi-liquid like state,
just like a superconductor can be viewed as a condensate of {\it Cooper pairs}
on top of a Fermi-liquid state \cite{normalstatenote}.  
To be more specific, the variational 
wave function of an AF insulator \ul{with \neel vector pointing in $\alpha$--direction}
is given by 
\begin{equation}
|\Psi_{AF} >\sim \prod_k (\tilde u_k + \tilde v_k c^\dagger_{k+Q} 
\sigma_\alpha c_k) |\Omega\rangle ,
\label{af}
\end{equation}
where $c^\dagger_k$ is a two component spinor creation opeator for an electron
state with wave vector $k$, $Q=(\pi,\pi)$ is the AF ordering vector, $\sigma_\alpha$
are the three Pauli spin matrices, $\tilde u_k$ and $\tilde v_k$ are the variational
parameters and $|\Omega\rangle$ is a Fermi-liquid like state at half-filling.
A magnon is defined by the operator $N_\alpha(k)=c^\dagger_{k+Q} 
\sigma_\alpha c_k$, and is formed by a triplet bound state of a particle and a hole pair
with respect to a half-filled state. 
On the other hand, a SC state is described by the following variational wave function
\begin{equation}
|\Psi_{SC} >\sim \prod_k ( u_k +  v_k c_k \sigma_y c_{-k})|\Omega\rangle ,
\label{sc}
\end{equation}
where $u_k$ and $v_k$ are the variational parameters for the SC state.
Here the Cooper pair is defined by the operator $B(k)=c_k \sigma_y c_{-k}$
and is formed by a singlet pair of holes with respect to a half-filled state.
We see that these two seemingly different states of matter are actually formally
equivalent, if one replaces the magnon operator $N_\alpha(k)$ 
by a Cooper pair operator $B(k)$.
This ``replacement" is exactly provided by the SO(5) rotations operator
\begin{equation}
\pi_\alpha = \sum_k g_k c_{k+Q} \sigma_\alpha \sigma_y c_{-k} \, , 
\label{pi}
\end{equation}
where the form factor $g_k=\sgn(\cos k_x-\cos k_y)$ is uniquely determined
by the group closure requirement \cite{henley,rabello}. 
This operator rotates the magnon and Cooper pair operators into each other
according to the following equation:
\begin{equation}
[\pi_\alpha, N_\beta(k)] = \delta_{\alpha\beta} g_k B(k) \ \ , \ \
[\pi^\dagger_\alpha, B(k)] = g_k N_\alpha(k)
\label{rotation}
\end{equation}
The internal wave functions of the elementary constituents of the
magnon and Cooper pairs are described the symmetries of 
$\tilde u_k \tilde v_k$ and $u_k v_k$ respectively. From
the above equation we see immediately that a $d=\cos k_x - \cos k_y$
form of the Cooper pair wave function will translate into 
a $|d| = |\cos k_x - \cos k_y|$ form of the magnon wave function.
This formal equivalence between a magnon condensate and a Cooper
pair condensate can be elevated to an SO(5) symmetry principle
whose dynamical consequence can be tested both numerically and
experimentally. Experimentally, the neutron resonance mode in
the SC state can be identified with the pseudo Goldstone modes
of the enlarged SO(5) symmetry group, which is a quantum fluctuation
from the SC state towards the AF state \cite{zhang,demler98}. 
This identification has recently been used to give an estimate for the 
condensation energy from the normal to the SC state \cite{demler98}.
Numerically, we have 
performed exact diagonalization studies of the Hubbard and the
$t-J$ models \cite{meixner,ederetal98}. In particular, for the $t-J$ model 
on a finite sized cluster, the ground
state at half-filling is a total spin singlet state. Low energy 
{\it bosonic} excitations above this ground state are either magnon or 
hole pair states. The relative energies of a single magnon and a
hole pair states can always be made degenerate by tuning the chemical
potential to a special value $\mu=\mu_c$. However, what is surprising 
is that the multi- magnon and hole pair states also become nearly
degenerate at this special value of the chemical potential \cite{ederetal98}. 
This is an indication that the interaction between the magnons
and the hole pairs are the nearly the same, originating from a
common type of physical interaction. Because of this near degeneracy of the
multi magnon and hole pair states, it is argued that the low energy {\it bosonic} 
excitations of the $t-J$ model near half-filling can be organized into
SO(5) symmetry multiplets \cite{ederetal98,ederetal99}.

The exact SO(5) symmetry \cite{rabello,henley} requires 
charge excitations at
half-filling to have the same gap as the collective spin-wave excitations. This condition is
violated in a Mott-Hubbard insulator, which has a large gap ($\sim eV$) to all charge
excitations while the spin excitations display no gap. 
In particular, in an exactly SO(5) symmetric description \cite{rabello,henley}, the
superconducting gap with its nodes would directly be mapped onto an AF gap, which then would
have precisely the same magnitude and would go go to zero at the nodes $(\pm \pi/2,\pm
\pi/2)$. Taking the Mott--Hubbard gap into account amounts to properly projecting out
these ''high-energy''
processes of order $\sim eV$ (Gutzwiller constraint) 
in the ''low-energy'' SO(5) rotation between AF and dSC states.
Recently, such a 
projection procedure has been carefully defined for 
the bosonic excitations and it was shown that the
Gutzwiller constraint can be implemented analytically and exactly within the
SO(5) theory \cite{zhanghu}. The central hypothesis of the
SO(5) theory states that the high $T_c$ superconductors can be described by
a fully SO(5) symmetric Hamiltonian supplemented by the Gutzwiller constraint.
The Gutzwiller constraint associated with the large on-site interaction reduces
the full SO(5) symmetry to a projected SO(5) symmetry with well defined
characteristics. 
The corresponding projected SO(5) model describes the low-energy bosonic degrees of freedom
near the AF/SC transition \cite{zhanghu}.

In this paper, we shall explore the consequence of the
projected SO(5) symmetry in the {\it fermionic} excitations of the AF and
SC states. In particular, we shall argue for the following two main points:
(i) The ARPES experiments show that the fermionic quasi-particles have 
a $s+|d|$ gap structure in the AF state. 
The projected SO(5) symmetry introduces the $s$ component of the AF gap
associated with the large on-site Coulomb energy, which is absent in the case of
pure SO(5) symmetry, and quantitatively relates the remaining $|d|$ component of the
AF gap with the $d$-wave SC gap.
\ul{While the $d$--wave SC gap is of the order of $J/10$,}
\ul{the obtained AF gap modulation is of the order of $J$.}
\ul{This is an unexpected consequence of the projection.} 
The correspondence between the $|d|$ component of the AF
gap structures and the $d$--wave SC gap structure is the signature of the projected
SO(5) symmetry in the fermionic excitation spectra.
(ii) The $t-J$ model with nearest neighbor hopping only gives an $s$ gap structure
in the AF state without the $|d|$ gap modulation. Therefore, while the {\it bosonic}
excitation spectra of the $t-J$ model can be organized by the projected SO(5)
symmetry, its {\it fermionic} quasi-particle spectra are {\it not} compatible with the
projected SO(5) symmetry and {\it not} compatible with the ARPES experiments. 
On the other hand, microscopically SO(5) symmetric
models with proper Gutzwiller projection can describe both the bosonic and the
fermionic excitation spectra of the high $T_c$ systems. 

Let us first see how a microscopically SO(5) symmetric Hamiltonian
with Gutzwiller projection can give rise to the $s+|d|$ gap structure as observed
in the ARPES experiments. This Hamiltonian contains the
manifestly SO(5)-symmetric terms, 
\begin{eqnarray}
H_{\kin}+H_{\INT} &=& 
\sum_{p,\sigma }\varepsilon _{p}c_{p,\sigma
}^{\dagger}c_{p,\sigma }+
V\sum_{\vec r_1,\vec r_2} \Big\{ 
\vec m(\vec r_1)\cdot 
\vec m(\vec r_2)+ 
\nonumber \\
& & \frac{1}{2}\left( \Delta(\vec r_1)\Delta(\vec r_2)^{\dagger}+
\Delta(\vec r_1)^{\dagger}\Delta(\vec r_2)\right) \Big\} .  \label{gl2}
\end{eqnarray}

Here, $H_{\kin}$ stands for the kinetic energy part with band dispersion 
$\varepsilon _{p}=-2t\left( \cos k_x+\cos k_y\right) $, valid for a
nearest-neighbor tight-binding model with hopping amplitude $t$ ($%
c_{p,\sigma }^{+}$ creates a hole with momentum $p$ and spin $\sigma )$.
Both $H_{\kin}$ and $H_{\INT}$ comprise the explicitly SO(5)-symmetric part
of the Hamiltonian \cite{rabello,henley}. 
$H_{\INT}$ contains a spin-spin interaction and a
pair-hopping term 
and , thus, serves as an example of the possible AF$\longleftrightarrow$dSC rotation, the
implications of which on the gap structure we shall investigate below.
$\vec m(\vec r_1)$ and $\Delta(\vec r_1)$ are N\'{e}el and $d-$wave
order parameters (operators) at site
$\vec r_1$.
The $d$-wave order parameter is built up of the usual superposition of nearest-neighbor (n.n.)
$d$-wave pairs \cite{noteex}. On the other hand, the \neel order parameter has an extended
internal structure \cite{noteintstruc}.
This extended internal structure 
in $\vec m(\vec r_1)$ is required by the SO(5)
symmetry, which may, at least in principle, be tested in experiments. 
In particular, it may be related effectively to spatially extended 
hoppings $\tp$ and $\tpp$ \cite{noteintstruc}.
The
essential physics of $H_{\INT}$ is that is has embedded, via the spin
interactions and pair hopping processes, the dynamical equivalence of spin
(triplet or magnon) excitations and pair excitations. 
We will argue below in more detail that this specific choice is physically motivated in that
it provides a possible AF$\longleftrightarrow $d-SC rotation, 
when the SC-part in $H_{\INT}$ corresponds to the
usual BCS reduced Hamiltonian for n.n. $d$-wave pairing. 
In particular, it
will be shown that 
the gap modulation is introduced in a rather natural
way already at 
the simplest mean-field level, i.e. in terms of BCS and
spin-density wave (SDW) gap equations (Hartree-Fock).

In addition to the fully SO(5) symmetric Hamiltonian, we need to implement
the Gutzwiller projection, which will reduce the full SO(5) symmetry to a
projected SO(5) symmetry. This can be implemented by the introduction
of a Hubbard $U$ interaction and by taking the limit of large $U$. 
Therefore, we arrive at the following Hamiltonian 
\begin{equation}
H=\left( H_{\kin}+H_{\INT}\right) +H_{U}+H_{\mu },  \label{g3}
\end{equation}
where 
\begin{equation}
H_{U}=U\sum_{i}n_{i\uparrow }n_{i\downarrow }.  \label{g4}
\end{equation}
$\left( n_{i\uparrow }=c_{i\uparrow }^{+}c_{i\downarrow }\right) $ is the
standard Hubbard interaction and $H_{\mu }=-\mu Q$ denotes the chemical
potential term. At this stage, a cautious reader may wonder about the problem
of double counting, since the perturbation theory of $H_{\kin}+H_U$ may
produce terms already included in $H_{\INT}$. But these pertubation effects
are of the order of $t^2/U$. Therefore, if we can properly take the 
$U\rightarrow\infty$ limit, we do not have to worry about the double
counting problem.  

\begin{figure}
\epsfxsize=8cm
\epsffile{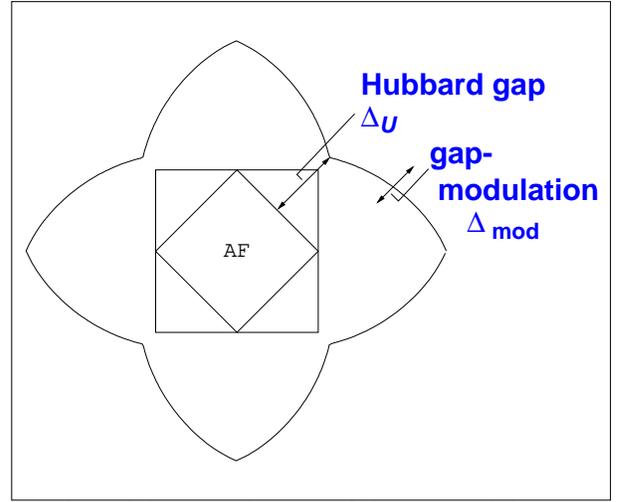}
\caption{Schematic drawing of the gap modulation and 
the constant gap in the antiferromagnetic phase as 
obtained from the Slave-Boson Mean-Field calculation. 
For graphical reasons we indicate the Hubbard gap $\Delta_U$
only as a relatively small offset 
to the modulation. In reality (and in our calculations) $\Delta_U \approx 4 t$ is about
one order of magnitude larger than the gap modulation. In the analog experimental 
plot in the inset of Figure 1 only the {\it difference} 
to the constant gap is shown and therefore no
offset is indicated.  }
\label{figgapaf}
\end{figure}

There are various ways to (approximately) study the Hamiltonian in Eq. (\ref{g3}).
Its physical content becomes transparent already on the simplest, i.e. Hartree-Fock
mean-field level. Earlier work by Schrieffer {et al.} 
on the Hubbard model \cite{schrieffer} shows that such a simple
mean field calculation can capture the basic physics in the 
strong-coupling limit. 
Consider first the SDW-type of solution for the \neel state. Here the gap
function $\Delta({\vec p})$ is connected to the SDW mean-field (polarized in
$z$-direction) by the standard relation:
\begin{equation}
\langle c_{\vec p + \vec Q,i}^\dagger \sigma^3_{ij} c_{\vec p,j}^{} \rangle 
= \frac{\Delta(\vec p)}{2 E(\vec p)} \, ,
\label{glorderparam}
\end{equation}
where, as usual, $E(\vec p) = \left( \eps^2(\vec p) + \Delta(\vec p) \right)^{1/2}$. When
introduced in equation (\ref{g3}) for the Hamiltonian, this mean-field order parameter
results in the self-consistency condition determining the gap $\Delta(\vec p)$,
\begin{equation}
\sum_{{\vec p}^\prime} V(\vec p, {\vec p}^\prime) \frac{\Delta({\vec p}^\prime)}
{2 E({\vec p}^\prime)} = \Delta(\vec p) \, .
\label{glselfcon}
\end{equation}
Here we shall take the factorized form of the SO(5) interaction, i.e. $V w_{\vec p} w_{\vec
p^\prime}$ \cite{noteex}, which also follows from equation (\ref{gl2}) and the
Fourier-transformed expression for the order parameter $\vec m(\vec r)$ in Ref.
\cite{noteintstruc}. Including the Hubbard $U$-term, $V(\vec p, {\vec p}^\prime)$ 
is given by
\begin{equation}
 V(\vec p, {\vec p}^\prime) = U + V w_{\vec p} w_{\vec p^\prime} \, ,
\label{glvppprime}
\end{equation}
It is now straightforward to show that the gap equation (\ref{glselfcon}) for the \neel
state as well as the corresponding BCS-type gap equation for the SC state yield the required
gap features:

First, note that the factorized form of the interaction $V(\vec p, {\vec p}^\prime)$ 
introduces
a similarly separable form of the AF gap, 
\begin{equation}
\Delta^{\af}(\vec p) = \Delta_U + \Delta_{\mod} w_{\vec p} \, ,
\label{glgapmodul}
\end{equation}
where, as before \cite{noteex}, 
$w_{\vec p} = | \cos p_x - \cos p_y |$. For large values of the Hubbard
interaction $\Delta_U$ is of the order of $U$. From 
experiment, $\Delta_{\mod}$ should be of
order $J$ \ul{while $\Delta_{\SC}$ should be of the order of $J/10$}
(this will be shown below 
to be the case 
using a Slave-Boson mean-field evaluation). 
Equation (\ref{glgapmodul})
then establishes the gap-modulation $\sim J |\cos p_x - \cos p_y |$ on top of a uniform gap
in the AF (Fig. \ref{figgapaf}).

Second, in formal analogy, in the $d$-wave SC state, the same gap equation (\ref{glselfcon})
holds, with three modifications: The first one is, of course, that $\Delta^{\SC}(\vec p)$
stands now for the $d$-wave SC order parameter with n.n. pairing \cite{noteex}. The second is
due to the well-known fact, that the $U$-term and, therefore, the constant gap term of the
AF drops out. This is because of the condition 
$\sum_{\vec p} U \frac{\Delta^{\SC}(\vec p)}{2 E(\vec p)} = 0$, which in real space is
nothing but the two pairing electrons avoiding the on-site Coulomb repulsion
\cite{scalapino95}.

\begin{figure}
\epsfxsize=8cm
\epsffile{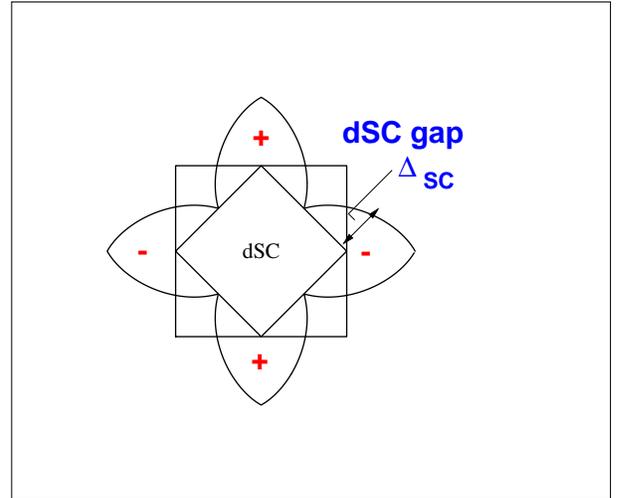}
\caption{Schematic drawing of the dSC-gap in the superconducting phase with 
conventions as in Figure 3 and the inset of Figure 1.
}
\label{figgapsc}
\end{figure}

Third, in the SC state, the relevant interaction in the gap equation is $V(\vec p, \vec
p^\prime) = V \, \left( w_{\vec p} \sgn(\cos p_x - \cos p_y) \right) 
\cdot \left( w_{\vec p^\prime}
\sgn(\cos p_x^\prime - \cos p_y^\prime) \right) $, 
finally  resulting in the gap function (Fig. \ref{figgapsc}),
\begin{equation}
\Delta^{\SC}(\vec p) = \Delta^{\SC} \cdot ( \cos p_x - \cos p_y) \, .
\label{glgapsc}
\end{equation}
Thus, both the AF gap in equation (\ref{glgapmodul}) and the SC gap have the required form.
What 
experiments tell us
is that while $\Delta^{\af}$ is of the order $J$,
$\Delta^{\SC}$ is an order of magnitude smaller. This requires a more involved mean-field
evaluation:
We have chosen a technique that is capable of dealing with the problem in a
way which is non-perturbative in the Hubbard interaction $U$. 
It is the Slave-Boson formalism introduced by Kotliar
and Ruckenstein \cite{kotliar}, \ul{which we treat by the usual saddle--point approximation.}
%

Confidence in this approximation derives from various observations: (i) 
the Slave-Boson mean-field approximation gives the
identical result as the so-called Gutzwiller approximation. The latter is
equivalent to the exact treatment of the Gutzwiller variational approach in
infinite dimensions 
\cite{kotliar,arrigoni}. 
We shall demonstrate that this
result is relevant for the ''projected'' SO(5) theory, i.e. for projecting
out the higher-energy states of the upper Hubbard band, as discussed above.
(ii) secondly, \ul{this approximation yields rather satisfying agreement with,} 
\ul{in principle exact, QMC simulations 
over a wide}
\ul{range of Coulomb correlations $U$ and values for the doping} \cite{lilly}. (iii)
the two main results of our study, i.e. that the AF gap modulation itself is
correlated to the SC gap by symmetry and that the Hubbard term $H_{U}$
induces the order of magnitude difference in the effects should be
independent of specific approximations.
This will be explicitely verified by comparing with the bare Hartree-Fock (HF) study.

\begin{figure}
\epsfxsize=8cm
\epsffile{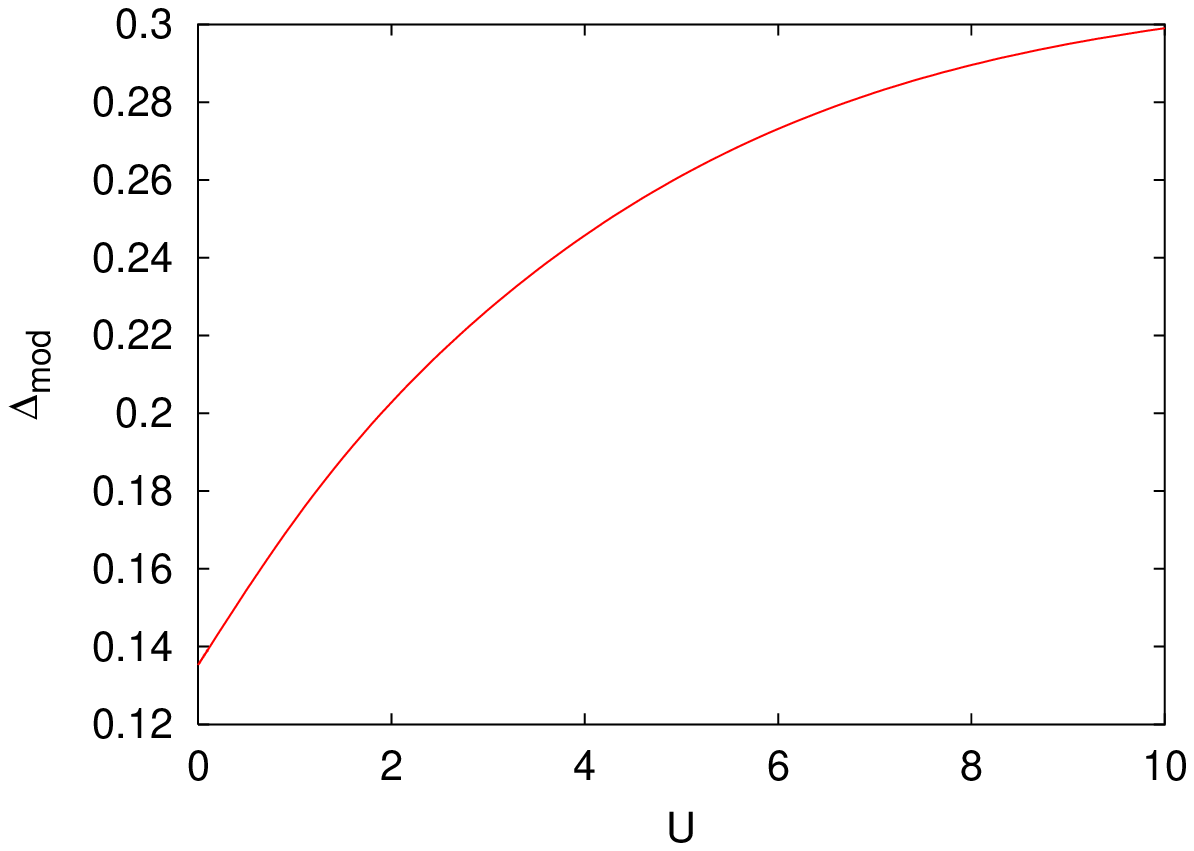}
\caption{The key result of the Slave-Boson Mean-Field calculation is the strong enhancement
of the gap modulation in the antiferromagnetic phase as $U$ is increased and double
occupancy becomes more and more forbidden. The symmetry-breaking $U$ is responsible for a
one order of magnitude difference of the gap modulation in the AF phase to the dSC gap in
the superconducting phase in the experimentally relevant region of $U\approx 8t$.
}
\label{figgapu}
\end{figure}

The results of the Slave-Boson evaluation of the Hamiltonian are presented
in terms of the gap features in Figs. \ref{figgapaf},\ref{figgapsc} and \ref{figgapu}.

Figs. \ref{figgapaf} and \ref{figgapsc} display the SC gap and the AF gap modulation again in the
same 2D version as in the experimental plot in Fig. \ref{figshen}. We note that 
as in HF, for symmetry reasons, 
the $d-$wave gap in the SC leads to a $|\cos
_{k_{x}}-\cos _{ky}|$-modulation of the AF gap. Thus, the AF gap is given by
$\Delta ^{\af}\left( \vec k\right) =\Delta _{U}+\Delta _{%
\mod}|\cos _{x}-\cos _{y}|$ displaying a cusp at the $\left(
k_{x},k_{y}\right) =\left( \pm \frac{\pi }{2},\pm \frac{\pi }{2}\right) $
wave vectors. The theory has only one free parameter, namely the SO(5)-%
coupling strength $V$ in Eq. (\ref{gl2}), which was chosen in such a way that it
gives a $d-$wave gap of the correct order of magnitude 
in the SC phase, i.e. $\Delta_{\SC}=0.02t\approxeq J/10$ \cite{notescgap}. 
The crucial observation in Figs. \ref{figgapaf},\ref{figgapsc} and \ref{figgapu} 
is then that, while the SO(5) interaction is responsible for the 
$d-$wave structure of both gaps,
it is a different mechanism, namely the Hubabrd gap, which is
responsible for the experimentally observed order of magnitude differences
in $\Delta_{\SC}\approxeq J/10$ and the AF gap modulation $\Delta _{\mod}\approxeq J$.

This is clearly demonstrated in Fig. \ref{figgapu}, which plots the amplitude $\Delta _{%
\mod}$ of the $d-$wave-like modulation as a function of the
Hubbard-gap $\Delta _{U}$ ($\Delta _{U}$ scales with $U$ for large $U$). We
note that increasing $U$ (and, thus, $\Delta _{U}$) and, therefore,
projecting out the doubly occupied states, strongly enhances the $d-$%
wave-like modulation in the AF gap. In particular, taking commonly accepted
values $U=8t$ and, therefore, $\Delta _{U}\approxeq 3t$ 
\cite{dagotto,preuss97} yields
an AF gap modulation of order $\Delta _{\mod}\approxeq
(0.25-0.3)t$. Thus, we find a radically different energy scale for $\Delta _{%
\mod}$ of order $J$ and $\Delta_{\SC}$ of order $J/10$, in agreement
with the ARPES data.

Let us now comment on the possible alternative explanations of the ARPES
data.  
Both Hubbard and $t-J$ Hamiltonians have been extensively studied in two
dimensions by employing exact diagonalization and Quantum-Monte-Carlo (QMC)
numerical techniques 
(see, for example, Refs.. 
\cite{dagotto,preuss97,eder97,duffy}
). 
There is
general consensus that both models have a serious deficiency, when compared
with the ARPES data in Fig. \ref{figshen}: they do not display the $d-$wave-type
dispersion, and the energies at $\left( \pi ,0\right) $ and $\left( \frac{%
\pi }{2},\frac{\pi }{2}\right) $ are essentially degenerate 
\cite{dagotto,preuss97,eder97}; 
as we have mentioned above, this is serious, since ARPES shows that
the evolution of the $\left( \pi ,0\right) $ feature is crucial to
understand the $d-$wave-like pseudo-gap. These numerical ''experiments''
have been further carried out with the inclusion of more-distant 
(next-nearest $\tp,\tpp$ etc.) 
neighbor hoppings 
\cite{kim,eder97,duffy}. 

However, this procedure requires delicate adjustments of the hopping integrals to
shift the structure near $(\pi,0)$ to the observed binding energy. 
In addition, at first
sight, it seems to be completely uncorrelated with two experimental findings which emphasize
the universal role of the magnetic energy scale $J$: 
one is the fact that the insulating bandwidth
itself scales with $J$ \cite{ronning}. 
The other is that, at present, it is not clear whether the
specific $\tp, \tpp$-choice is simultaneously in accordance with the
universal picture which recently emerged from the incommensurate spin fluctuations of two
major classes of \htc materials, \lasrco and \ybco \cite{aepplikastner,mook}. 
We have already indicated (see Ref. \cite{noteex}) that $\tp,\tpp$-hoppings may arise
effectively from the SO(5)-part of the Hamiltonian in Eq. (\ref{gl2}), namely from the
spin interaction term 
$V\sum_{\vec r_1,\vec r_2} \vec m(\vec r_1)\cdot \vec m(\vec r_2) $.
However, in addition to $\tp, \tpp$ this interaction effectively produces longer-range hoppings
(although with magnitude decreasing with distance) which are essential in order to produce
the cusp-like feature of the AF-gap. Indeed, a finite-range hopping (i.e. limited to $\tp,
\tpp$) is not sufficient to produce a cusp-like feature \cite{notemod}.

Summarizing, our results show that the recent experimental discovery of the 
$s+|d|$ gap structure in the AF state can be naturally explained by the
concept of projected SO(5) symmetry, which relates the $|d|$ gap modulation
of the AF state with the $d$-wave gap of the SC state.
If taken as an exact SO(5)
theory, without the physically relevant symmetry-breaking term $H_{U}$, this
would have resulted in an exact mapping, where the AF gap has nodes. It has
been shown here that, while $H_{U}$ breaks the symmetry and restores the
Mott-Hubbard gap, it does not break the experimentally observed correlation
between the $d-$wave gap features. In fact, as demonstrated here, $H_{U}$ is
pivotal in explaining the order of magnitude differences 
\ul{between the superconducting gap $\Delta_{\SC}$ } \ul{and the $d$--wave--like 
modulation of the AF gap $\Delta _{\mod}$.}
Just like the neutron resonance mode can be interpreted
as the reflection of AF correlation in the SC state, the ARPES experiment
can be interpreted as the reflection of the SC correlation in the AF state.
Our results also show that the Hubbard and $t-J$ models still ''miss a piece'',
while a microscopically SO(5) symmetric Hamiltonian with proper Gutzwiller
projection can correctly describe both the bosonic and the fermionic excitation
spectra of the high $T_c$ superconductors. 

We thank R.B.\ Laughlin, D.J.\ Scalapino, R.\ Eder, and O.K.\ Andersen for helpful
discussions.
W.H. and E.A.
would also like to acknowledge the support and hospitality of the Stanford Physics
Department, where most of this work was carried out. This work was  supported
from the DFG (AR 324/1-1 and HA 1537/17-1), DFN (contract No. TK598-VA/D3),
BMBF (05SB8WWA1) and NSF (grant number DMR-9814289).


\newpage

\narrowtext

\begin{figure}
\epsfxsize=8cm
\epsffile{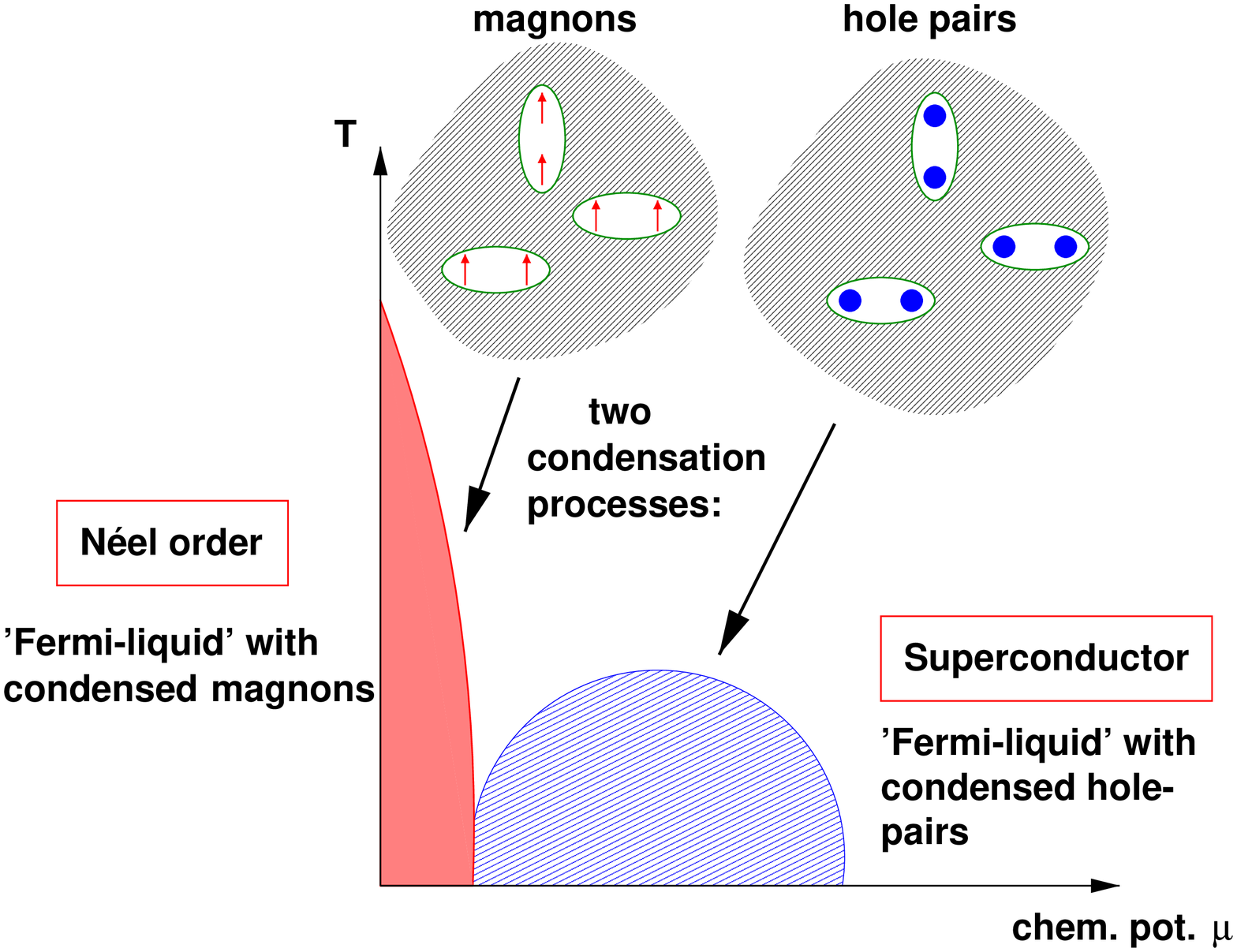}
\caption{The $(T,\mu)$-phase diagram for \htc superconductors reveals the intimate
relation between the AF and dSC state: The AF state can be viewed as a condensation of
magnons (triplet excitations) on top of a Fermi-liquid like state, just as the
superconducting state can be represented as the condensation of hole pairs. SO(5) symmetry
relates the magnon onto the hole-pair states. The ARPES experiments probe the internal
structure of a Cooper pair and magnon, respectively, thereby revealing the correlation
between antiferromagnetism and superconductivity.}
\label{figphase}
\end{figure}

\newpage

\newpage

\end{multicols}


\begin{thebibliography}{99}
\bibitem{notedisp}
Notice that the ARPES
data measure only the dispersion of the first ionization states, i.e. the
''topmost'' valence band dispersion. In case of electron-hole symmetry, as
in our theory, this dispersion is then translated into an equivalent $|\cos
k_{x}-\cos k_{y}|$-modulation of the AF gap.
\bibitem{normalstatenote}
While this ''normal state'' may have anomalous properties it has the same symmetry as the
Fermi-liquid state (see Ref. \cite{eder98}). We identify the normal state as a Fermi-liquid
in the sense that it is adiabatically continuable to the over-doped metallic state.
Alternatively to this momentum--space description, one may start from a more localized
real--space version:
above $T_{\mathrm{\neel}}$ up to temperatures of order $J \sim 1000 K$, a variety of experiments
suggest that the charge carriers have a tendency to bind into spin--singlet states. However,
the  spin singlets are spatially disordered. They form a
spin--liquid state $|\tilde \Omega \rangle$, which we may take to be of short--range RVB
(resonating--valence--bond) form, where the singlets are formed by nearest--neighbor spins
\cite{ederetal99,eder98}. Consider then two ''fixed'' neighboring sites in the AF , say in
the spin configuration $(\up \down)$. This ''fixed'' configuration can be generated
by superposition of the singlet $(\up\down) - (\down\up)$ with the $S_z =0$ triplet
$(\up\down) + (\down\up)$. To obtain the macroscopic 2D antiferromagnet, we have to admix
into the spin--liquid at high temperatures a macroscopic number of triplets $t_z(\vec Q)$.
The AF ordered state can then be generated forming the coherent state
$|\psi \rangle \sim e^{t_z(\vec Q)} |\tilde \Omega \rangle$, i.e. via a ''condensation'' of
triplets into the lowest--energy state \cite{ederetal99,eder98}. 
However, in the SO(5) theory, the z-component triplet with 
momentum $(\pi,\pi)$ and the hole pair with momentum $(0,0)$
 are dynamically equivalent, in the
same sense as e.g. proton and neutron are considered dynamically equivalent in the isospin
theory of nuclear physics. The ''RVB type'' of spin liquid can thus
serve as an alternative common reference state $|\tilde \Omega\rangle$ to consider
the condensation of magnons and charged Bosons (hole pairs) on equal footing, thus revealing
the analogy between AF and d-SC states.
The detailed relation between the Fermi--liquid momentum--space description and the
real--space spin liquid picture is not clear yet and presently under intensive study.
\bibitem{noteex}
The SC part of $H_{\INT}$ is, in momentum space, $2 V \{ \Delta \Delta^\dagger +
\Delta^\dagger \Delta \}$, where $\Delta$ is the usual n.n. $d$-wave order parameter, i.e.
$\Delta = \sum_{p,\sigma} \frac{1}{2} ( \cos p_x - \cos p_y ) c_{\vec p,\up}^\dagger
c_{-\vec p,\down}^{}$. This ''separable'' form in $\Delta$ and $\Delta^\dagger$ fixes the
general SO(5)-invariant interaction with the coupling $V(\vec p, \vec p^\prime; \vec q)$ to
be separable in momentum space and given by $V(\vec p, \vec p^\prime; \vec q) = V
\delta_{\vec q,\vec Q} w_{\vec p} w_{\vec p^\prime}$, where $w_{\vec p} = | \cos p_x - \cos
p_y|$ (see also Ref. \cite{rabello}).
\bibitem{noteintstruc}
The extended internal structure of $\vec m(\vec r_1)$ [ i.e. $\vec m(\vec r_1) = \sum_{\vec
r_2,i,j} w(\vec r_1 - \vec r_2) e^{i \vec Q - \vec r_2} c^\dagger_{\vec r_1,i} \vec
\sigma_{ij} c^{}_{\vec r_2,j}$, where $\vec \sigma$ are the usual $2 \times 2$ Pauli
matrices] introduces on a mean field level (replacing $\vec m(\vec r_2)$ in equation
(\ref{gl2}) by its expectation value $\langle \vec m(\vec r_2) \rangle$) longer-ranged
hopping terms $\tp$ (next n.n.) and $\tpp$ (third n.n.). Similarly to the 
commonly used 
spatially extended hoppings \cite{kim,eder97,duffy}, their signs alternate. This
stems from the fact that $w(\vec r)$ is the Fourier-transform of $w(\vec p) = |\cos p_x -
\cos p_y |$ (see Ref. \cite{noteex}) and 
takes the values ($\vec r = (m,n)$ on the 2D
lattice): $w(m,n)=0$, if $m+n=$ odd, $w(0,0)=\frac{8}{\pi^2}$, $w(1,1)=-\frac{8}{3\pi^2}$
and $w(2,0) = \frac{8}{9 \pi^2}$, etc. This results in $\tp \sim - 4 V w(0,0) w(1,1) \approx
-O(J)$ and $\tpp \sim + 4 V w(0,0) w(2,0)$ with $\tpp/\tp = - 1/3$, where we 
used $V=0.05 t$ (see Ref. \cite{notescgap}).
\bibitem{notescgap}  For the SC gap (around optimal doping) experiments give a $d$-%
wave gap magnitude of order $J/10\sim 10-20meV$ [ ]. A reasonable value for $%
J$ is $J=0.2t$. For the choice $V=0.05t$, we obtain - for the commonly
accepted value of the Hubbard correlation $U=8t$ - at 10 \% doping $\Delta
_{SC}=0.02t\approxeq J/10$.
\end{thebibliography}
\end{document}